\documentclass[a4paper]{jpconf}
\usepackage[T2A]{fontenc}
\usepackage[utf8]{inputenc}
\usepackage[english]{babel}
\usepackage{graphicx}
\usepackage{nicefrac}
\usepackage{amsmath}
\usepackage{amsfonts}
\usepackage{amssymb}
\usepackage{amsthm}
\usepackage{epsf}
\usepackage{bm}
\usepackage{longtable}
\LTcapwidth=\textwidth
\usepackage{caption}
\AtBeginDocument{%
\renewcommand{\tablename}{\textbf{Table}}
}
\newcommand{\gd}{g_\mathrm{D}^{}}
\newcommand{\dgint}{\Delta g_{\mathrm{int}}}
\newcommand{\dgqed}{\Delta g_{\mathrm{QED}}^{}}
\newcommand{\dgrec}{\Delta g_{\mathrm{rec}}}
\newcommand{\dgns}{\Delta g_{\mathrm{NS}}^{}}
\newcommand{\dgintlo}{\dgint^{(0)}}
\newcommand{\dgintz}{\dgint^{(1)}}
\newcommand{\dgintzz}{\dgint^{(2)}}
\newcommand{\dgintzqed}{\dgint^{(1)}[\mathrm{QED}]}
\newcommand{\dgintps}{\dgint^{(1)}[+]}
\newcommand{\dgintns}{\dgint^{(1)}[-]}
\newcommand{\dgqedol}{\Delta g_{\mathrm{QED}}^{(1)}}
\newcommand{\dgqedtl}{\Delta g_{\mathrm{QED}}^{(2)}}
\newcommand{\dgse}{\Delta g_{\mathrm{SE}}^{}}

\newcommand{\dgvp}{\Delta g_{\mathrm{VP}}^{}}
\newcommand{\aZ}{\alpha Z}
\newcommand{\Rnucl}{R_\mathrm{nucl}}
\begin{document}

\title{Ground-state \emph{g} factor of middle-$Z$ boronlike ions}

\author{V~A~Agababaev$^{1,2}$, D~A~Glazov$^{1}$, A~V~Volotka$^{1,3}$, D~V~Zinenko$^{1}$, V~M~Shabaev$^{1}$ and G~Plunien$^{4}$}

\address{$^1$ Department of Physics, Saint-Petersburg State University, 199034 Saint-Petersburg, Russia}
\address{$^2$ Saint-Petersburg State Electrotechnical University ``LETI'', 197376 Saint-Petersburg, Russia}
\address{$^3$ Helmholtz-Institut Jena, D-07743 Jena, Germany}
\address{$^4$ Institut f\"ur Theoretische Physik, Technische Universit\"at Dresden, D-01062 Dresden, Germany}

\begin{abstract}
Theoretical calculations of the interelectronic-interaction and QED corrections to the \emph{g} factor of the ground state of boronlike ions are presented. The first-order interelectronic-interaction and the self-energy corrections are evaluated within the rigorous QED approach in the effective screening potential. The second-order interelectronic interaction is considered within the Breit approximation. The nuclear recoil effect is also taken into account. The results for the ground-state \emph{g} factor of boronlike ions in the range $Z$=10--20 are presented and compared to the previous calculations.
\end{abstract}
%
%
The past two decades have been marked by intensive development of the \emph{g}-factor studies in highly charged ions \cite{sturm:17:a,shabaev:15:jpcrd}. Experimental precision has reached the level of $10^{-9}$--$10^{-11}$ for hydrogenlike and lithiumlike ions \cite{haefner:00:prl,verdu:04:prl,sturm:11:prl,sturm:13:pra,wagner:13:prl}. Cooperative experimental and theoretical work led to the most accurate up-to-date value of the electron mass \cite{sturm:14:n}. The most stringent test of the many-electron QED effects in the presence of magnetic field has been achieved with middle-$Z$ lithiumlike ions \cite{wagner:13:prl,volotka:14:prl,koehler:16:nc,yerokhin:17:pra}. Simultaneous high-precision \emph{g}-factor measurement for two calcium isotopes \cite{koehler:16:nc} and the rigorous evaluation of the relativistic nuclear-recoil effect \cite{shabaev:17:prl,malyshev:17:jetpl} have opened perspective for testing bound-state QED effects beyond the Furry picture (external field approximation for the nucleus). Independent determination of the fine structure constant $\alpha$ is possible in \emph{g}-factor studies with high-$Z$ \cite{shabaev:06:prl} or middle-$Z$ \cite{yerokhin:16:prl} boronlike, lithiumlike and hydrogenlike ions. The ARTEMIS experiment presently implemented at GSI aims at measurement of the \emph{g} factors of both ground $[(1s)^2(2s)^2 2p]~{}^2P_{1/2}$ and first excited $[(1s)^2(2s)^2 2p]~{}^2P_{3/2}$ states of boronlike argon \cite{lindenfels:13:pra}. In this regard, the leading interelectronic-interaction, QED and recoil corrections to these \emph{g} factors were calculated in Refs.~\cite{glazov:13:ps,shchepetnov:15:jpcs} employing the bound-state QED perturbation theory and the configuration-interaction Dirac-Fock-Sturm (CI-DFS) method. In Ref.~\cite{verdebout:14:adndt} the GRASP2K program package based on the relativistic multi-configuration Dirac-Hartree-Fock method was used to obtain the energy levels, the hyperfine interaction constants and the \emph{g} factors in beryllium-, boron-, carbon- and nitrogen-like ions in the range $Z$=8--42. In Ref.~\cite{marques:16:pra} the \emph{g} factors of boronlike ions in the range $Z$=14--92 were evaluated within the multi-configuration Dirac-Fock method using the MCDFGME code. Significant difference between the results of Refs. \cite{glazov:13:ps,shchepetnov:15:jpcs,verdebout:14:adndt,marques:16:pra} motivated us to perform independent calculations within the framework of the bound-state QED perturbation theory. In this paper, we present the results for the ground-state \emph{g} factor of boronlike ions in the range $Z$=10--20.
The relativistic units ($\hbar = c = 1$) and the Heaviside charge unit ($\alpha=e^2/(4\pi), e<0$) are used throughout the paper. 

The total \emph{g}-factor value of boronlike ion with spinless nucleus can be written as
\begin{equation}
  g = \gd + \dgint + \dgqed + \dgrec + \dgns
\,,
\end{equation}
where the leading contribution can be found analytically from the Dirac equation with the point-nucleus potential,
\begin{equation}
  \gd = \frac{2}{3}\bigg[\sqrt{2\big(1+\sqrt{1 - (\aZ)^2\big)}} - 1 \bigg] = \frac{2}{3} - \frac{1}{6}(\aZ)^2 - \dots
\,,
\end{equation}
and $\dgint$, $\dgqed$, $\dgrec$ and $\dgns$ denote the interelectronic-interaction, QED, nuclear recoil and nuclear size corrections, respectively.

The correction due to the interelectronic interaction is considered within the perturbation theory. The term of the first order in $1/Z$ is calculated within the rigorous QED approach, i.e., to all orders in $\aZ$. The second-order contribution is considered within the Breit approximation. In Refs.~\cite{volotka:14:prl,volotka:12:prl} the two-photon-exchange corrections to the \emph{g} factor and to the hyperfine splitting have been evaluated within the rigorous QED approach for lithiumlike ions. The formulae presented in Ref.~\cite{volotka:12:prl} can be used to derive the corresponding expressions within the Breit approximation. A distinctive feature of the \emph{g}-factor calculations is the necessity to account for the negative-energy-states contribution, since it is comparable in magnitude to the positive-energy counterpart.

In order to account approximately for the higher-order corrections, an effective screening potential is introduced in the Dirac equation. It leads to emergence of the zeroth-order contribution --- difference between the \emph{g}-factor values for the effective screening and the pure Coulomb potentials. The corresponding counterterms have to be taken into account in the first- and second-order contributions. We consider four different screening potentials --- core-Hartree (CH), Dirac-Hartree (DH), Kohn-Sham (KS) and Dirac-Slater (DS). Explicit formulae for these potentials can be found e.g. in Refs.~\cite{sapirstein:02:pra,glazov:06:pla}. We note that the evaluation of the two-photon-exchange contribution in the pure Coulomb nuclear potential is related to some numerical problems in case of the boronlike ions.

In table~\ref{tab:g-int} the breakdown of the interelectronic-interaction correction is given in terms of the \emph{g}-factor contributions multiplied by $10^6$. The first-order term is split into three parts: the positive-energy-states ($\dgintps$) and negative-energy-states ($\dgintns$) contributions and the QED contribution ($\dgintzqed$). The two former are obtained within the Breit approximation. The latter is found as the difference between the rigorous QED result and the Breit-approximation result. The total value of $\dgint$ is found as a sum of the evaluated contributions,
\begin{equation}
  \dgint = \dgintlo + \dgintz + \dgintzz
\,,
\end{equation}
where
\begin{equation}
  \dgintz = \dgintps + \dgintns + \dgintzqed
\,.
\end{equation}
We choose the result for the Kohn-Sham potential as the final one. The total value of $\dgint$ would not depend on the effective potential, if all orders of the perturbation theory were taken into account rigorously. Thus the spread of the results for different potentials can serve as an estimation of the uncertainty due to the unknown higher-order contributions. As one can see from the table, the maximal difference of the values of $\dgint$ varies between $1.6\times 10^{-6}$ for $Z$=10 and $0.7\times 10^{-6}$ for $Z$=20. Interelectronic-interaction corrections of the third and higher orders have been evaluated for lithiumlike ions within the CI-DFS \cite{volotka:14:prl} and CI \cite{yerokhin:17:pra} methods. The results obtained in these papers suggest that this estimation of the uncertainty is quite reliable. We can also estimate the unknown QED part of the two-photon-exchange correction $\dgintzz$ as not more than $0.2\times 10^{-6}$ based on the results of Ref.~\cite{volotka:14:prl}. 

One-loop QED correction $\dgqedol$ is given by the sum of the self-energy and the vacuum-polarization contributions,
\begin{equation}
  \dgqedol = \dgse + \dgvp
\,.
\end{equation}
The self-energy correction for the $2p_j$ states was calculated to all orders in $\aZ$ in Ref.~\cite{yerokhin:10:pra}. The numerical approach was based on the Dirac-Coulomb Green's function in order to achieve rather high accuracy, which is especially difficult for low nuclear charge. Instead, we use the approach developed in Refs.~\cite{glazov:06:pla,volotka:06:epjd}, which is based on the DKB finite basis set~\cite{shabaev:04:prl}. Although, it is generally less accurate, it allows one to easily incorporate arbitrary spherically symmetric binding potential. In order to account approximately for the many-electron QED effects we use effective screening potentials, the same ones that we use for evaluation of $\dgint$: core-Hartree, Dirac-Hartree, Kohn-Sham and Dirac-Slater. The results of the calculations are given in table~\ref{tab:g-qed}.

The one-electron vacuum-polarization correction $\dgvp$ is negligible for the $2p_{1/2}$ state in the considered range of $Z$. The dominant effect of the vacuum polarization arises from the two-electron diagrams, where the $1s$ and $2s$ electrons of the closed shells come into play. Still, it is much smaller than the total theoretical uncertainty: for $Z$=18 it was estimated as $6.4\times 10^{-9}$ in Ref.~\cite{glazov:13:ps}. The two-loop QED contributions $\dgqedtl$ are taken into account to the zeroth order in $\aZ$ according to Ref.~\cite{grotch:73:pra}.

The nuclear-recoil contribution was calculated for boronlike argon in Ref.~\cite{glazov:13:ps} including the leading relativistic corrections and the screening effect. In Ref.~\cite{shchepetnov:15:jpcs} the first-order interelectronic-interaction correction was considered using the nonrelativistic approximation for the recoil operator. Recently, the nuclear recoil effect to the \emph{g} factor of boronlike ions has been evaluated with the relativistic recoil operator in the zeroth and first orders in $1/Z$ \cite{glazov:18:os}. These results are used in the present compilation.
The finite-nuclear-size correction $\dgns$ for $2p_{1/2}$ state to the leading order in $\aZ$ can be written as \cite{glazov:02:pla}
\begin{equation}
\label{eq:dgns}
  \dgns = \frac{(\aZ)^6}{16} m_e^2 \Rnucl^2
\,,
\end{equation}
where $\Rnucl$ is the nuclear root-mean-square radius. For $Z$=10--20 equation~(\ref{eq:dgns}) gives the values of the order $10^{-13}$--$10^{-11}$, i.e., much smaller than the total theoretical uncertainty.

In table~\ref{tab:g} we present the individual contributions and the total values of the \emph{g} factor of boronlike ions in the range $Z$=10--20. The Kohn-Sham values of $\dgint$ (see table~\ref{tab:g-int}) and $\dgqedol$ (see table~\ref{tab:g-qed}) are employed. Despite the different approach to evaluation of the second- and higher-order interelectronic-interaction effects, our results for argon are in agreement with Ref.~\cite{shchepetnov:15:jpcs}. For comparison we present also the data from Ref.~\cite{verdebout:14:adndt} and Ref.~\cite{marques:16:pra}. One can see that the difference between the values of Verdebout \etal and of the present work grows monotonically from $0.000\,045$ for $Z$=10 to $0.000\,088$ for $Z$=20. The corresponding difference with the values of Marques \etal ranges from $0.000\,187$ for $Z$=14 to $0.000\,283$ for $Z$=20. At present, we can not clearly identify the source of this disagreement. However, we suppose that the contribution of the negative-energy states was not completely taken into account in Refs.~\cite{verdebout:14:adndt,marques:16:pra}. 

We note also that the nonlinear contributions in magnetic field are important in boronlike ions \cite{lindenfels:13:pra,glazov:13:ps}. Recently, the second- and third-order effects have been evaluated within the fully relativistic approach for the wide range of $Z$ \cite{varentsova:18:pra}. While the second-order effect is not observable in the ground-state Zeeman splitting, the third-order effect has to be taken into account. Its relative contribution amounts to $2.6\times 10^{-8}$ for $Z$=10 and $3.5\times 10^{-11}$ for $Z$=20 at the field strength of 1 Tesla and it scales as $B^2$.

In conclusion, the \emph{g} factor of boronlike ions in the range $Z$=10--20 has been evaluated with an uncertainty on the level of $10^{-6}$. The leading interelectronic-interaction and QED effects have been calculated to all orders in $\aZ$. The higher-order interelectronic-interaction and nuclear-recoil effects have been taken into account within the Breit approximation.
%
%
%
%
%
%
%
%
%
%
%
\begin{longtable}{lrrrr}
\caption{\label{tab:g-int}
Interelectronic-interaction correction to the \emph{g} factor of boronlike ions in terms of $\Delta g\times 10^{6}$. The contributions of the zeroth ($\dgintlo$), first ($\dgintz$) and second ($\dgintzz$) orders of the perturbation theory obtained with the core-Hartree (CH), Dirac-Hartree (DH), Kohn-Sham (KS) and Dirac-Slater (DS) screening potentials. The first-order term is split into the contributions of the positive-energy ($\dgintps$) and negative-energy ($\dgintns$) spectra calculated within the Breit approximation and the QED part ($\dgintzqed$).}
\\
\hline
& \multicolumn{1}{c}{CH} 
& \multicolumn{1}{c}{DH}
& \multicolumn{1}{c}{KS}
& \multicolumn{1}{c}{DS}
\\
\hline
\multicolumn{5}{c}{$Z = 10$}\\
\hline
$\dgintlo$   &   379.092  &   470.808  &   390.491  &   345.422  \\
$\dgintps$   & $-$30.899  & $-$91.371  & $-$39.453  &  $-$6.168  \\
$\dgintns$   &  $-$1.820  & $-$38.936  &  $-$4.897  &    15.864  \\
$\dgintzqed$ &  $-$0.148  &  $-$0.118  &  $-$0.149  &  $-$0.166  \\
$\dgintzz$   &    10.139  &    14.568  &    10.531  &     0.003  \\
$\dgint$     &   356.364  &   354.951  &   356.523  &   354.956  \\
\hline
\multicolumn{5}{c}{$Z = 12$}\\
\hline
$\dgintlo$   &   461.050  &   578.458  &   474.753  &   418.092  \\
$\dgintps$   & $-$38.052  &$-$111.255  & $-$47.404  &  $-$8.707  \\
$\dgintns$   &  $-$3.321  & $-$53.249  &  $-$7.622  &    18.987  \\
$\dgintzqed$ &  $-$0.291  &  $-$0.245  &  $-$0.294  &  $-$0.320  \\
$\dgintzz$   &    10.119  &    14.802  &    10.141  &     0.343  \\
$\dgint$     &   429.505  &   428.509  &   429.573  &   428.395  \\
\hline
\multicolumn{5}{c}{$Z = 14$}\\
\hline
$\dgintlo$   &   543.283  &   686.232  &   559.217  &   490.972  \\
$\dgintps$   & $-$45.268  &$-$131.208  & $-$55.478  & $-$11.194  \\
$\dgintns$   &  $-$4.725  & $-$67.395  & $-$10.212  &    22.206  \\
$\dgintzqed$ &  $-$0.506  &  $-$0.440  &  $-$0.509  &  $-$0.546  \\
$\dgintzz$   &    10.104  &    14.963  &     9.899  &     0.519  \\
$\dgint$     &   502.888  &   502.151  &   502.915  &   501.957  \\
\hline
\multicolumn{5}{c}{$Z = 16$}\\
\hline
$\dgintlo$   &   625.826  &   794.263  &   643.939  &   564.093  \\
$\dgintps$   & $-$52.499  &$-$151.206  & $-$63.582  & $-$13.599  \\
$\dgintns$   &  $-$6.043  & $-$81.416  & $-$12.694  &    25.515  \\
$\dgintzqed$ &  $-$0.802  &  $-$0.713  &  $-$0.809  &  $-$0.857  \\
$\dgintzz$   &    10.088  &    15.083  &     9.718  &     0.601  \\
$\dgint$     &   576.570  &   576.011  &   576.572  &   575.752  \\
\hline
\pagebreak
\hline
\multicolumn{5}{c}{$Z = 18$}\\
\hline
$\dgintlo$   &   708.721  &   902.650  &   728.969  &   637.488  \\
$\dgintps$   & $-$59.722  &$-$171.243  & $-$71.670  & $-$15.902  \\
$\dgintns$   &  $-$7.275  & $-$95.330  & $-$15.078  &    28.914  \\
$\dgintzqed$ &  $-$1.194  &  $-$1.080  &  $-$1.204  &  $-$1.266  \\
$\dgintzz$   &    10.068  &    15.180  &     9.566  &     0.622  \\
$\dgint$     &   650.598  &   650.177  &   650.584  &   649.855  \\
\hline
\multicolumn{5}{c}{$Z = 20$}\\
\hline
$\dgintlo$   &   792.014  &  1011.470  &   814.355  &   711.194  \\
$\dgintps$   & $-$66.922  &$-$191.319  & $-$79.712  & $-$18.082  \\
$\dgintns$   &  $-$8.417  &$-$109.147  & $-$17.365  &    32.408  \\
$\dgintzqed$ &  $-$1.695  &  $-$1.552  &  $-$1.708  &  $-$1.785  \\
$\dgintzz$   &    10.043  &    15.261  &     9.429  &     0.597  \\
$\dgint$     &   725.023  &   724.714  &   724.998  &   724.332  \\
\hline
\end{longtable}
%
%
\begin{table}[b]
\caption{\label{tab:g-qed}
Self-energy correction $\dgse$ to the \emph{g} factor of boronlike ions obtained with the core-Hartree (CH), Dirac-Hartree (DH), Kohn-Sham (KS) and Dirac-Slater (DS) screening potentials in terms of $\Delta g\times 10^{6}$.}
\vspace{0.5cm}
\begin{center}
\begin{tabular}{lrrrr}
\hline
$Z$
& \multicolumn{1}{c}{CH} 
& \multicolumn{1}{c}{DH}
& \multicolumn{1}{c}{KS}
& \multicolumn{1}{c}{DS}
\\
\hline
$10$  & $-$773.05  & $-$773.06  & $-$772.99  & $-$772.95  \\
$12$  & $-$772.43  & $-$772.49  & $-$772.36  & $-$772.29  \\
$14$  & $-$771.61  & $-$771.70  & $-$771.53  & $-$771.44  \\
$16$  & $-$770.60  & $-$770.71  & $-$770.50  & $-$770.39  \\
$18$  & $-$769.39  & $-$769.51  & $-$769.26  & $-$769.13  \\
$20$  & $-$767.95  & $-$768.10  & $-$767.81  & $-$767.65  \\
\hline
\end{tabular}
\end{center}
\end{table}
\newpage
%
%
\begin{table}
\caption{\label{tab:g}
Ground-state \emph{g} factor of boronlike ions in the range $Z$=10--20. The values obtained with the Kohn-Sham potential are used for the interelectronic-interaction correction $\dgint$ (see table~\ref{tab:g-int}) and the one-loop QED correction $\dgqedol$ (see table~\ref{tab:g-qed}). The \emph{g}-factor values from Refs.~\cite{shchepetnov:15:jpcs,verdebout:14:adndt,marques:16:pra} are given for comparison.}
\vspace{0.5cm}
\begin{center}
\begin{tabular}{lr@{}lr@{}l}
\hline
\\
& \multicolumn{2}{c}{${}^{20}_{10}$Ne$^{5+}$}
& \multicolumn{2}{c}{${}^{24}_{12}$Mg$^{7+}$}
\\[6pt]
\hline
Dirac value $\gd$                     &    0.&665\,777\,663     &    0.&665\,385\,559     \\
Interelectronic interaction $\dgint$  &    0.&000\,356\,5\,(16) &    0.&000\,429\,6\,(12) \\
One-loop QED $\dgqedol$               & $-$0.&000\,773\,0\,(4)  & $-$0.&000\,772\,4\,(5)  \\
Two-loop QED $\dgqedtl$               &    0.&000\,001\,2       &    0.&000\,001\,2       \\
Nuclear recoil $\dgrec$               & $-$0.&000\,015\,2\,(12) & $-$0.&000\,013\,6\,(7)  \\
\hline
Total value $g$                       &    0.&665\,347\,2\,(20) &    0.&665\,030\,4\,(15) \\
\hline
Total value $g$ \cite{verdebout:14:adndt}
                                      &    0.&665\,392          &    0.&665\,084          \\
\hline
\\
& \multicolumn{2}{c}{${}^{28}_{14}$Si$^{9+}$} 
& \multicolumn{2}{c}{${}^{32}_{16}$S$^{11+}$}
\\[6pt]
\hline
Dirac value $\gd$                     &    0.&664\,921\,417     &    0.&664\,384\,860     \\
Interelectronic interaction $\dgint$  &    0.&000\,502\,9\,(10) &    0.&000\,576\,6\,(8)  \\
One-loop QED $\dgqedol$               & $-$0.&000\,771\,5\,(6)  & $-$0.&000\,770\,5\,(8)  \\
Two-loop QED $\dgqedtl$               &    0.&000\,001\,2       &    0.&000\,001\,2       \\
Nuclear recoil $\dgrec$               & $-$0.&000\,012\,3\,(4)  & $-$0.&000\,011\,1\,(3)  \\
\hline
Total value $g$                       &    0.&664\,641\,7\,(12) &    0.&664\,181\,1\,(12) \\
\hline
Total value $g$ \cite{verdebout:14:adndt}
                                      &    0.&664\,704          &    0.&664\,252          \\
Total value $g$ \cite{marques:16:pra}
                                      &    0.&664\,829\,(40)    &    0.&664\,400\,(46)    \\
\hline
\\
& \multicolumn{2}{c}{${}^{40}_{18}$Ar$^{13+}$}
& \multicolumn{2}{c}{${}^{40}_{20}$Ca$^{15+}$}
\\[6pt]
\hline
Dirac value $\gd$                     &    0.&663\,775\,447     &    0.&663\,092\,678     \\
Interelectronic interaction $\dgint$  &    0.&000\,650\,6\,(7)  &    0.&000\,725\,0\,(7)  \\
One-loop QED $\dgqedol$               & $-$0.&000\,769\,3\,(9)  & $-$0.&000\,767\,8\,(10) \\
Two-loop QED $\dgqedtl$               &    0.&000\,001\,2\,(1)  &    0.&000\,001\,2\,(1)  \\
Nuclear recoil $\dgrec$               & $-$0.&000\,009\,1\,(2)  & $-$0.&000\,009\,3\,(2)  \\
\hline
Total value $g$                       &    0.&663\,648\,8\,(12) &    0.&663\,041\,8\,(12) \\
\hline
Total value $g$ \cite{verdebout:14:adndt}
                                      &    0.&663\,728          &    0.&663\,130          \\
Total value $g$ \cite{marques:16:pra}
                                      &    0.&663\,899\,(2)     &    0.&663\,325\,(56)    \\
Total value $g$ \cite{shchepetnov:15:jpcs}
                                      &    0.&663\,647\,7\,(7)  &      &                  \\
\hline
\end{tabular}
\end{center}
\end{table}
%
%
%
\ack
The work was supported in part by RFBR (Grant No.~16-02-00334), by DFG (Grant No.~VO 1707/1-3), by SPbSU-DFG (Grant No.~11.65.41.2017 and No.~STO 346/5-1) and by SPbSU (Grant No.~11.40.538.2017). V.A.A. acknowledges the support by the German-Russian Interdisciplinary Science Center (G-RISC). The numerical computations were performed at the St. Petersburg State University Computing Center.
%
%
\newpage
%
%
\section*{References}
%
%
\bibliography{database}
%
%
%
\end{document}